# Extension of the non-parametric cluster-based time-frequency statistics to the full time windows and to single condition tests

Christian Beste [1]*, Daniel Kaping [2,1]*, Tzvetomir Tzvetanov[**][3,4]

1 Cognitive Neurophysiology, Department of Child and Adolescent Psychiatry, Faculty of Medicine, TU Dresden, Germany
2 National Institute of Mental Health, Klecany, Czech Republic
3 CAS Key Laboratory of Brain Function and Disease, School of Life Sciences, University of Science and Technology of China, 230027, Hefei, Anhui, People's Republic of China
4 Anhui Key Laboratory for Affective Computing and Advanced Intelligent Machines and School of Computer and Information, Hefei University of Technology, 230009 Hefei, Anhui, People's Republic of China

\* equally contributed
\*\* corresponding author

Address for correspondence
Tzvetomir Tzvetanov
School for Computer and Information, Hefei University of Technology
230009 Hefei, Anhui, People's Republic of China
tzvetan@hfut.edu.cn

**Abstract**
Oscillatory processes are central for the understanding of the neural bases of cognition and behaviour. To analyse these processes, time-frequency (TF) decomposition methods are applied and non-parametric cluster-based statistical procedure are used for comparing two or more conditions. While this combination is a powerful method, it has two drawbacks. One the unreliable estimation of signals outside the cone-of-influence and the second relates to the length of the time frequency window used for the analysis. Both impose constrains on the non-parametric statistical procedure for inferring an effect in the TF domain. Here we extend the method to reliably infer oscillatory differences within the full TF map and to test single conditions. We show that it can be applied in small time windows irrespective of the cone-of-influence and we further develop its application to single-condition case for testing the hypothesis of the presence or not of time-varying signals. We present tests of this new method on real EEG and behavioural data and show that its sensitivity to single-condition tests is at least as good as classic Fourier analysis. Statistical inference in the full TF map is available and efficient in detecting differences between conditions as well as the presence of time-varying signal in single condition.

## 1. Introduction and test extension to the full- time-frequency window

Research in neuroscience analyses oscillations in neural activity and behaviour across a broad range of frequencies. Synchronization and locking of oscillatory processes are suggested to be central to the neural basis of cognition (Buzsáki, 2006; Cavanagh and Frank, 2014; Fries, 2015; Siegel et al., 2012; Singer, 2011). In parallel, studies have shown that oscillatory measures are predictive of behavioural processes, spanning from long time-scale chronobiological effects to the short time-scales of visual attention processes (Fiebelkorn et al., 2013; Landau and Fries, 2012; VanRullen, 2016). In such studies, typically non-parametric cluster-based permutation tests (Maris & Oostenveld, 2007; Maris, 2012) are applied for testing differences in time evolution of frequency activity. In this note, we argue about the applicability of the test to the full time-frequency (TF) window, independent of the analysis time window size and wavelet sizes, thus allowing the researchers to infer differences in TF content up to the borders of the time window.

To examine the evolution of the frequency content in time series, time-frequency (TF) analyses are conducted using wavelet transformations (Cohen, 2014; Mallat, 2009). The TF representation of the signal has the advantage that it can disassociate changes at different frequencies across time and does not require stationarity of the recorded signal (Başar et al., 2001; Quiroga et al., 2001). The TF representation of a data set is extracted either from wavelet sparse decomposition, representing the signal in a minimalistic space of orthogonal wavelets, or more generally by obtaining the signal's convolution with a predefined set of wavelets. The latter approach is of interest here and the results are interpreted as the amplitude and phase of a given frequency at each time point of the signal. It is a 2 dimensional function which for ease of visualization is generally plotted in 2D color-coded format where, for example, the amplitude (or its square the power) is color-coded, and vertical and horizontal axes represent frequency and time, respectively (refer Figure 1).

Such 2D TF maps are extracted for different experimental conditions and these two or more conditions are compared through two a-priori possible strategies. The TF window can be identified using either a hypothesis-driven or a data-driven strategy (Cohen, 2014; Dippel et al., 2017, 2016; Mückschel et al., 2016). When there are no clear a-priori assumptions on the TF window to be analysed, a data-driven strategy needs to be employed. This involves non-parametric cluster-based permutation testing that is done in two stages. Stage-1 selects bin values in the TF map that have a given amount of significant difference; it is done by testing the null hypothesis that the two compared conditions' bin values are equal (e.g. simple t-tests). Then, Stage-2 tests the null hypothesis that the size of the cluster (or some other variable of the cluster) of contiguous significant bins in the TF map is a random realization of the measurement-analysis method (Maris, 2012; Maris and Oostenveld, 2007), i.e. it is statistically probable to obtain such a value for the cluster found in the data. In this process, corrections for multiple comparisons are taken into account by using the cluster-based statistics of Stage-2.

However, the above procedure is applied only on TF results obtained within the cone-of-influence and thus contains rarely explicitly stated problems. The first issue is the difficulty to infer the signal's TF content outside the cone-of-influence (COI; see further below, and ch.6, pp.215-218, (Mallat, 2009)). The second problem is the length of the chosen analysis time window of the data, possibly creating unwanted data overlap and overlaps in wavelet-signals. These two points are not dissociable in any TF analysis, since fixing the data time window and the wavelet size completely defines the cone-of-influence domain. The COI designates the window of validity within the time domain of the convolution operation where the wavelet is fully contained in the time window of the data, and thus allows amplitude/phase comparisons across time points.

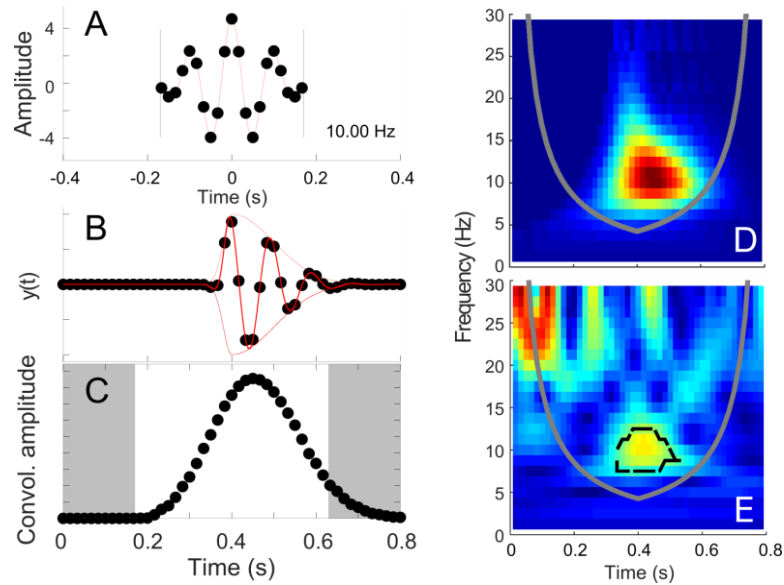

**Figure 1**: Example of wavelet analysis of a signal giving its 2D time-frequency representation, and the appearance of the cone-of-influence. (A) An example of a wavelet with a central frequency of 10 Hz and its width in time (vertical black thin lines; imaginary sine component not shown). (B) An example of a pure transient signal, y(t), of frequency 10 Hz. The thick red line shows the theoretical signal, the thin red line its amplitude evolution. (C) The result of convolving the 10 Hz wavelet in (A) and the signal in (B) (phase not shown); grey shaded areas correspond to the time locations of the wavelet at which it is not fully contained in the time window. Repeating the convolution of signal (B) with wavelets of different frequencies gives the 2D result depicted as a colour map in (D). The area between the grey cone represents the cone-of-influence domain where the results of the convolution operations are comparable to each other (white area in (C)). (E) shows the same as (D) but y(t) is the sum of the signal in (B) with a white Gaussian noise of mean zero and standard deviation equal to the amplitude of the signal in (B); example TF map of one single simulation with signal's amplitude bump visually present; dashed black lines delineate the cluster showing a tendency of an effect (pcluster=0.087). Wherever present, dots depict the time discrete version of the theoretical function (60 Hz sampling rate).

An example is shown in Figure 1. If the time window of the data is 0.8 seconds (refer Figure 1), a wavelet extending 0.34 seconds (peak frequency at 10 Hz) will provide amplitudes and phases that can be compared to each other only within a time window of 0.46 seconds (i.e. 0.8 minus 0.34) (white area in Fig.1C). The size of the time-frequency window, where inferences about signal's content can be made, is thus dependent on wavelet's frequency, which gives the $1/t$ structure of the COI (see Fig.1D). Consequently, it is contentious what can be inferred outside the COI domain (Mallat, 2009), and a statistical method that allows to make inference about the data within the full TF window is needed. This becomes all the more important when one considers to decrease the size of the time window, for either decreasing measurement time, especially important in studies of measures of oscillations in behavioural parameters (e.g. perceptual oscillations: (Fiebelkorn et al., 2013; Landau and Fries, 2012; VanRullen, 2016), or avoiding overlaps in signals close together.

The main interest implementing a TF analysis is to demonstrate that a tested condition differs from a control condition; i.e. to show that the two compared conditions show statistical differences in TF maps within the chosen TF window. This is accomplished in two stages (Maris, 2012; Maris and Oostenveld, 2007): *Stage-1* tests independently for each bin in the TF map whether the two measures could have been obtained from the same reference distribution and all bins are marked that are significant at a predefined critical level ($\alpha_{bin}$). Then, *Stage-2* creates clusters of contiguous significant bins and compares the data cluster(s) to a reference distribution of the cluster(s). This is achieved using a Monte-Carlo randomization procedure of the original data (permutation test). This overall procedure can be called a non-parametric "TF bump test", that is, it allows to detect the presence of at least one amplitude-bump, or power-bump, or "phase"-bump, in an otherwise similar distribution of TF waves between the two conditions. For example, taking a hand made example of two conditions, (a) noise only and (b) noise with added signal of amplitude equal to one standard deviation of the white Gaussian noise (Fig.1B,E), the test detects the presence of some (normally unknown) frequency signal within the time window of measurement (Fig.1E shows one "nearly-detected" result). The important feature in this procedure is that it dissociates the TF variable of interest, e.g. amplitude, from the variable used to infer the presence of an effect in the TF space. In this regard *Stage-2* is independent from the exact variable obtained in *Stage-1*. Therefore, the fact that outside of the cone-of-influence the convolution operation on the data gives amplitude/phase parameters that are not comparable with the values inside the COI is not relevant any-more. This results from *Stage-1* comparing each bin's data differences to its own reference distribution, obtained from exactly the same mathematical operation and procedure, which is similarly biased across all repetitive computations of the reference values in the bin. Consequently, the non-parametric cluster-based test performed on the bin-wise statistical TF map can be used in the full TF extent up to the borders of the time window chosen by the experimentalist. In doing so, one still needs to carefully consider the time extent of the wavelets since they symmetrically pool measures across time, but this method allows to infer the presence of effects in much smaller time windows than before.

Here we illustrate how the test works in the full TF map defined by the experimentalist. A simple case is presented where the test is applied on EEG data (refer methods section). The event-related potentials (ERPs) contain two separate conditions, shown as red (condition *X*) and black (condition *Y*) curves in Figure 2A together with their variability. Based on a visual inspection, the two ERPs look very similar, despite some differences at the beginning of the epoch.

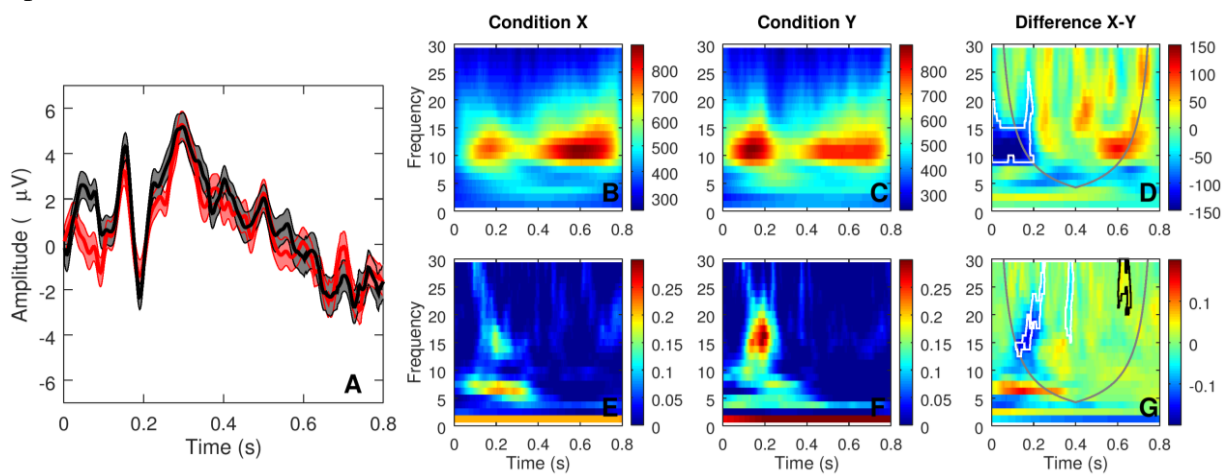

**Figure 2**: Example of TF analysis in short time window applied to EEG data. (A) Mean event related potentials (ERPs) at electrode Oz for two conditions X (red) and Y (black) as a function of time after cue onset (time point zero). Shaded areas show the s.e.m. (n=97). (B,C) Amplitudes of the TF decomposition for conditions X and Y showing the alpha band (10-12 Hz) activity in both cases. (D) Difference of amplitudes map and the significant cluster. Grey solid lines depict the COI. (E-F) same as (B-D) but for variable PPC (see text). (white solid contours=clusters of significant negative differences; black solid contours= clusters of significant positive differences).

In the TF map of the amplitudes (Figure 2B,C), one can see that both conditions have a strong alpha band (~10-12 Hz) activity with two amplitude peaks (bumps), the first around 150 ms and the second between 400 to 800 ms. The two bumps seem to show different amplitudes between the two conditions. When computing the difference map and the statistics between the two conditions, only one strong cluster centred on 100 ms and 12 Hz is identified as significant (see Figure 2D). A big portion of this cluster is found outside the COI and thus would have been missed due to this short time window of analysis.

The complementary variable of PPC (pairwise phase consistency; (Vinck et al., 2010)), which describes the phase consistency of the signal at a given frequency and time across repeated measures, was also analysed (Figure 2E-G). It provided a different picture than the amplitude. Two strongly phase locked signals appeared around 200 ms, one around ~6 Hz and one around 15 Hz (Fig.2E,F) with the 15 Hz having stronger PPC values in condition *Y* than *X* and vice versa for the ~6 Hz case. Applying the permutation statistics showed significant clusters at three locations, at the visualised bump around {t=200ms, f=15 Hz} location, but also a high frequency (15-30 Hz) localized at ~400 ms synchronisation and a small increase in PPC in condition *X* compared to *Y* at frequencies above 20 Hz at ~600 ms. The later two effects correspond to very weak phase synchronisations (below 0.1).

From the above clarifications and example of application, it is apparent that the non-parametric cluster-based statistical test can be applied within the full TF window chosen by the researcher. Therefore, this method can be applied irrespective of the cone-of-influence, including all TF bins up to the border of the time window of measurement. This is important for cognitive neuroscience research to link electrophysiological and behavioural data, as it permits the analysis of small time-frequency windows for comparisons between EEG and behavioural data.

## 2. Limits of the non-parametric cluster-based statistical test

Maris and Oostenveld (2007) have described the essential conditions for performing the cluster-based permutation test. One important assumption is the equality of the probability distributions which are obtained from the data in the two compared conditions (section 4.2, (Maris and Oostenveld, 2007). This is an important limitation of the method as shown below:

Let us assume that one wants to compare two measures $y_1(t)$ and $y_2(t)$ which are the sum of signals $s_{1,2}(t)$ and a noise term $\varepsilon(t)$ that affects both measures similarly:

$$\begin{aligned} y_1(t) &= s_1(t) + \epsilon_1(t) \\ y_2(t) &= s_2(t) + \epsilon_2(t) \end{aligned} \quad \text{(Equ. 3)}$$

Let us further assume that we want to test whether there are differences in TF amplitudes between the two measures, irrespective of the phase of the signals, and that there are *n* repeated measurements performed.

The null hypothesis states that:

$$\forall (t,f): |y_1^*(t,f)| - |y_2^*(t,f)| = |s_1^*(t,f) + \epsilon_1^*(t,f)| - |s_2^*(t,f) + \epsilon_2^*(t,f)| = 0, \quad \text{(Equ. 4)}$$

where * indicates the complex TF representation of the measures obtained from the convolution with the complex wavelets, and operator |.| indicates the magnitude of the complex number. The permutation test assumes the equality of the probability distributions $|y_{1,2}^*(t,f)|$. Since the noise must be of the same structure between the two measures this also implies similarity of the two signals probability distributions $|s_{1,2}^*(t,f)|$. This assumption allows to exchange any of the *n* values of $|y_1^*(t,f)|$ with its equivalent measures $|y_2^*(t,f)|$. By computing the difference of these two permuted values one creates a reference distribution by successive random re-sampling. However, and importantly, if the probability distributions of the two signals differ sufficiently strong from each other in comparison to the noise probability distributions, then the assumption will not be valid anymore and the test is also not valid. This can easily be seen in equation (4) and one simple example: If one takes the noise as null, or sufficiently small in comparison to the signal of interest, and $s_2(t)=0$, then the method becomes a test of presence of a signal $s_1(t)$. However, performing the permutation of the two measures across data points creates a reference distribution that is a mixture of a Dirac distribution at zero (a single peak distribution) and the signal $s_1(t)$'s distribution. Thus, existing test can only be performed, if the two signals' probability distributions are similar, or if the signals' probability distributions are not too different from the noise probability distributions. The important point is that the full TF map of the data should not be systematically different from the mean TF map obtained from the randomization procedure.

### 3. Extension for testing single condition

Based on the above arguments of similarity of the two distributions that are compared, it turns out that the method is also applicable to a single condition, e.g. does $y_1$ have an oscillatory effect in it's time window (independently of $y_2$). Such an idea was already tested previously (Fiebelkorn et al., 2013; Landau and Fries, 2012), but it was limited to classic Fourier analysis with simple test at each frequency with the necessary multiple comparisons adjustments. Based on the argumentations of the two previous sections it is therefore also possible to apply the non-parametric cluster-based test to single condition measures in TF space. If there is only a single condition, the reference TF distribution, to which the data TF map is compared, must be obtained from a different permutation. Given that the only available information is the raw data $y_1(t)$, we can create control data sets by randomly permuting in the time dimension the original data ; i.e. the null hypothesis states that the data is pure noise, and thus data points at all times are equivalently interchangeable. Thus, one can compare the TF map of the data to the reference TF map obtained from TF analysis of the simulated measures $y_{1,p}(t)$ (subscript p indicates the p-th permutated/shuffled data set; see Methods for details). The resulting reference TF map and its distribution at each bin can be used to perform the test at *Stage-1* to examine which bins in the data are significantly different from a noise TF map. The single bin's distributions that are obtained are not necessarily following well behaved Gaussian like distributions, therefore care must be taken to check the bin's distributions. Instead of transforming to z-values, one can also directly use the $p_{bin}$ values obtained at *Stage-1* that are distribution independent. The final cluster-based test remains as previously described in *Stage-2*.

In order to know whether the reference distribution is valid, we can check, as described in the previous section, whether the reference TF map is globally similar to the data TF map. This provides a simple check that one can easily perform and visualize by plotting the data TF map, the mean Monte-Carlo TF map, and the proportion of significant bins found at *Stage-1*. The test is implemented as follows: (1) perform *Stage-1* analysis of significant bins in the full TF map; (2) compute the number of significant bins across the TF map and divide by the total number of bins in the map to obtain $P_{sign}$; this last number represents the proportion of bins with significant effects, and respectively the proportion of bins equal to the reference distribution ($P_{equ}=1-P_{sign}$). The value of $P_{sign}$ can be used as a variable for testing whether the randomization procedure created a proper reference TF map. In case of no signal its value should be around $\alpha_{bin}$ and in case of small signal effect its value should stay in a relatively low range.

However, a few remarks are necessary. This last variable provides a decision variable ($P_{sign}$) for estimating how realistic is the reference TF map data given the data TF distribution, and does not say anything about specific localized effects in the TF domain. It naturally incorporates the time-frequency discretization chosen by the researcher and it is dependent on the relative size of the expected/observed significant cluster to the TF map size. For example, if the signal-to-noise (SNR) is very high, then $P_{sign}$ will be the proportion of area of your significant cluster in the full TF map, and will thus give an idea on the extent of the effect (seen in the 2D map). If $P_{sign}$ is too high, say above 10%, and by inspection of your data's TF map the cluster size you expected is not so big, then the randomization procedure did not create a proper reference distribution because too many significant "noise" bins were present. The above description explains why the single-condition test is appropriate only for detecting the presence of relatively small effects in noisy conditions or localized oscillatory effects in low noise data. For example, this single-condition test was applied on the two cases of strong SNR (Figure 1B and D) and low SNR (Figure 1E, SNR=1). In the case of strong SNR the cluster is highly significant (probability to obtain such a cluster by chance,: $p_{cluster}$=0.0043; cluster size=100 bins, Figure 1D) while for SNR of 1 the cluster delineating the signal bump shows only a tendency ($p_{cluster}$=0.087, cluster size=55 bins, Fig.1E; a second cluster is located in the top-left corner and is not significant, $p_{cluster}$=0.25).

It is important to note that this single condition test has very different application rules than the non-parametric cluster-based test proposed by Maris & Oostenveld (2007), and which was essentially created for comparing the final data of two measured conditions (ERPs, TF maps etc.). Here the two distributions of test and reference, which are tested against each other, are obtained after applying a transformation, here TF decomposition, on the data and its shuffled version. In the original procedure the permutation was between the two compared conditions while in this single condition test the permutation procedure is carried on the raw data not the compared conditions. Thus, care must be taken to ascertain the similarity of the two distributions at *Stage-1* that are used for carrying the cluster-based permutation test at *Stage-2*. The previous two paragraphs provide a simple description of how to check the similarity of the two distributions before application of the cluster-based test.

Last, the above single-condition test and its limitations should extend to comparisons in other domains as for example 1D time domain (e.g. Maris & Oostenveld, 2007, Figure 1) or classic 1D Fourier analysis (Landau & Fries, 2012; Fiebelkorn et al., 2013). Furthermore, the test keeps the family-wise error rate (FWER) at the appropriate level since this is based on *Stage-2*, the cluster-based statistics (Maris & Oostenveld, 2007), while the above single-condition validity check is concerned with *Stage-1* distributions.

## 4 Application of the full TF window test on a single condition

For demonstrating the applicability of the test on a single condition, we use data from an exogenous attention effect on discrimination performance of target stimulus (see Methods for details).

The target stimulus could appear randomly within a 800 ms window after cue onset and we measured the reaction time (RT) of the subjects for giving a correct response. The mean RTs of 22 subjects are plotted on Figure 3A and we observe the typical decrease in RTs with longer cue to target onset asynchrony (Posner, 1980). Out of the very low frequency component at 1-2Hz, performing a classic Fourier analysis showed no specific frequency content (Figure 3B). Applying the single condition test on the TF map showed that, across subjects, in the data there are two strong positive clusters of amplitudes around 2-4 Hz and 15-21 Hz that start at time zero (Figure 3C), but also that there seem to be less fluctuations than expected in two other locations (white clusters in Figure 3C).

This example also helps to demonstrate the sensitivity of this single condition test despite the known asymmetric noise distribution of the RTs (Luce, 1986; Ratcliff, 1993). This particularity of RTs noise distribution creates the oscillatory fluctuation in amplitudes observed further away from the COI (Figure 3C) that is clearly visible in the mean reference distribution of TF amplitudes (Figure 3D). Despite these known effects (Mallat, 2009), because of the dissociation of the cluster-based test and the physical variable, the test remains valid and is insensitive to these amplitude oscillations outside the COI.

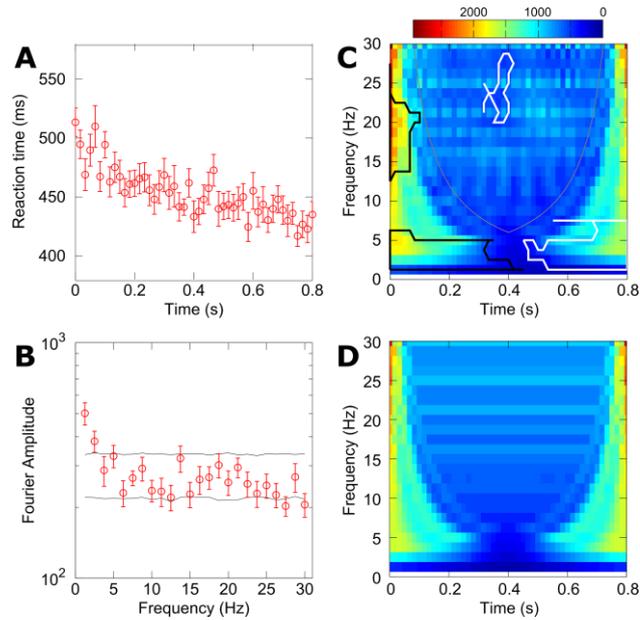

**Figure 3**: Example of TF analysis in short time window applied on behavioural data of a single condition. (A) Mean reaction times (RTs) for correct responses as a function of cue-to-target once asynchrony (n=22 subjects; positive times are target appearing after cue). (B) Mean Fourier amplitudes of the RTs in (A). Error bars are s.e.m. (C) Mean TF amplitudes obtained from the RTs in (A) and the significant clusters of stronger amplitudes than expected (black solid lines) and weaker amplitudes than expected (white solid lines). (D) Mean TF amplitudes of the reference distributions obtained by the single-condition test that shows the oscillations of the amplitudes outside the COI (see text). Here $\sigma_G=1.2/f_G$ was used.

The above example gave a case where the single-condition test seemed successfully applied. Now, we provide also an example that demonstrates when the test can be easily discarded because of a non-valid reference distribution. For that purpose, we use one of the EEG data set from the previous section (red curve reploted in Fig.4A). Applying *Stage-1* analysis of the single condition test gave a proportion of bins $P_{sign}$=0.797 in the data TF map that are significantly above the reference TF map. If one further carries the cluster-based test the resulting positive and negative significant clusters turn out to be very large, together almost covering the entire TF map (Figure 4B). In this extreme example the EEG data cannot be tested with the single condition test because the reference distribution of the individual bins substantially deviated from the data TF map. Two reasons of the test failure in this particular case are due to a combination of data pre-processing and the peculiarly strong signal around 10-12 Hz in the tested TF map combined to very low ERP noise.

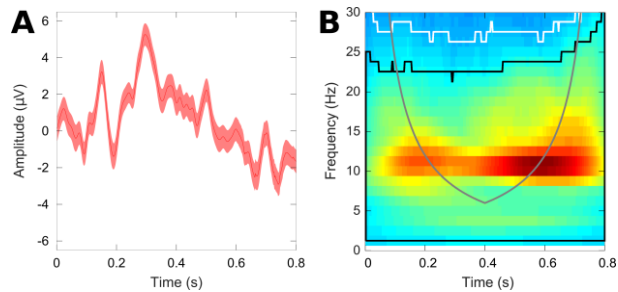

**Figure 4**: Example of failure applying the single condition test. (A) one ERP signal from figure 2A. (B) Applying the single condition test created a reference TF distribution at each bin that made the data TF map systematically above the reference distributions in ~79.7% of the map (black contour) and below the reference distributions in ~11.9% of the map (white contour).

## 5. Sensitivity of the single condition test

Because the application of the method to a single-condition seems not yet documented, we further analyzed the sensitivity of the method in comparison to the classic Fourier analysis usually applied to such small window sizes (Fiebelkorn et al., 2013; Landau and Fries, 2012). For that purpose simulations were carried out to investigate the sensitivity of the method for detecting a theoretical signal embedded in noise at various signal-to-noise (SNR) ratios and as a function of the distance of signal's peak from the border of the time window. First, we asked how the test performs on classic reaction time data that have a typical asymmetric distribution. When the signal to be detected had a main frequency of 5 Hz, the single-condition test performed globally as good as, or better than, classic Fourier analysis (Figure 4A; differences above 0.138 for Percent Detect around 0.5 between Fourier and cluster-based permutation test are significant at 5% double-sided test, see Zar, 1999). With higher signal frequencies, its sensitivity at detecting a signal embedded in noise was better than the classic Fourier analysis (Figure 4B and C). Second, these simulations were repeated with a normal white Gaussian noise. The results also showed that the single-condition non-parametric test outperformed the classic Fourier analysis (Figure 4D to F).

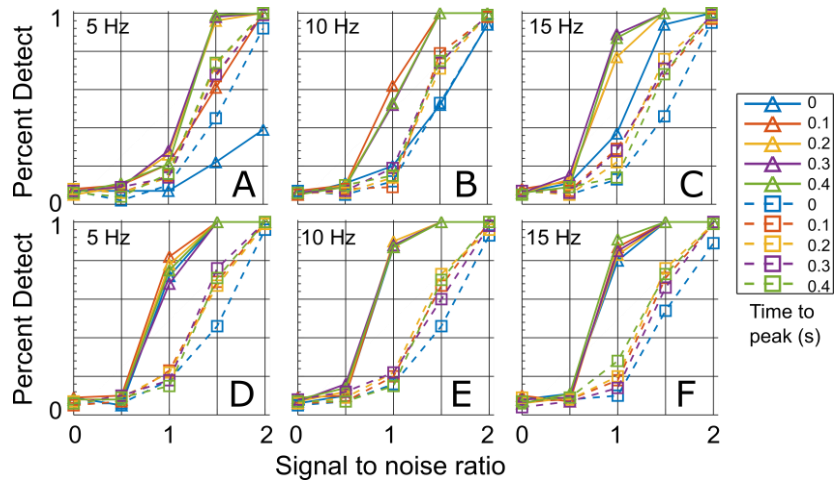

**Figure 5**: Detection performance of the single-condition non-parametric cluster-based test (triangles) and Fourier analysis (squares) methods for a simulated signal embedded in log-normal noise (A-B) or white Gaussian noise (D-F). Three possible signal frequencies were simulated, of 5, 10 and 15 Hz. Each estimation/symbol is obtained from 100 simulations. Rightmost panel displays the legend for signal's peak possible location from the border (0) up to the middle (0.4) of the time window.

## 6. Relation to other single-condition tests

It should be noted that in another scientific field researchers have proposed single condition tests that are closely related to the one we described above. In analysing time series of non-linear dynamic systems, people are often confronted with a measurement of a single condition (Bradley and Kantz, 2015; Kantz and Schreiber, 1997). From the beginning the question appeared of how to assess that the data is obtained from a non-linear dynamic system instead from a more classical linear (but stochastic) dynamics. The proposed solution, dubbed "surrogate data/surrogate time series", was to create Monte-Carlo samples that preserve some property of the original data, e.g. probability distribution of the data across time measures or additionally the amplitude of Fourier spectrum, but not its time evolution, i.e. the time points or phase can be randomized (Schreiber and Schmitz, 2000) and then to asses, with the use of a non-parametric distribution of the chosen variable of interest, obtained from the Monte-Carlo randomization procedure, whether the data could be a simple random realization of the measurement-analysis method given the hypothesis that it is a linear system.

This statistical procedure is very similar to the procedure we described for the single condition test. The only difference is that in the TF analysis proposed in this work the hypothesis tested is whether the data is simply a sample of random stochastic measures and does not contain any specific time varying signal superimposed to the noise. Thus, the cross-check that we proposed, whether the test is applicable (see section 3), seems to be a different instantiation of the method of surrogate data analysis where the researcher defines a-priori what component of the data and surrogates must be identical.

**7. Summary**

In summary, we provide advancements in the use of time frequency statistical inference analysis. First, we unveil that the problem of statistical inference outside the cone-of-influence is solved by the use of the cluster-based statistical procedure, which extends the statistical inference to the full time-frequency window used in the analysis. Second, we demonstrated that this novel method can be applied to short analysis time windows. This allows for TF analysis of behavioural data in short-time series which grants the advantage of close integration of different sorts of data used in cognitive neuroscience. Third, the non-parametric method can be applied to single conditions for detecting the presence of time-varying signal. Its sensitivity to detect the presence of a signal is as good, or better than, than classic Fourier analysis.

## 8. Materials and Methods

*Real EEG data*

The tutorial EEG dataset provided was continuously recorded and amplified using a QuickAmp system (Brain Products, Inc.) with 60 Ag-AgCl electrodes placed at standard scalp positions. The prospective study employed a visual cued stimulus discrimination task (controlled via Psychophysics toolbox (www.psychtoolbox.org (Brainard, 1997; Pelli, 1997) of which two conditions of 97 trials each have been selected from a single subject at electrode Oz. The dataset was processed through custom written Matlab scripts (The Mathworks Inc.) using the Fieldtrip toolbox (Oostenveld et al., 2011) and EEGLAB signal processing environment (Delorme and Makeig, 2004). After a band-pass filter ranging from 0.5 to 35 Hz was applied to the dataset, irregular technical and movement artefacts like grimacing, yawning, sneezing etc. were removed by means of a manual raw data inspection. Then, an independent component analysis (ICA, infomax algorithm) was applied to discard recurring physiological artefact like eye blinks, horizontal and vertical eye movements, as well as pulse artefacts. Components reflecting these artefacts were discarded before the EEG was reconstructed. This data pre-processing step was followed by the data segmentation step. The data was segmented at the onset of the cue stimulus and the segment ended 800ms after the cue onset. Within this time window, the target stimulus to discriminate appeared randomly. The Matlab/Octave based functions demonstrating the application of the method on the EEG data set can be downloaded at http://vision.ustc.edu.cn/packages/TutorialDataSetFunctions_TFanalysis.zip.

*Behavioural data*

The reaction time data are a subset of the full data, corresponding to one condition. Subjects had to discriminate the orientation of a small orientated stimulus (~0.7 degrees diameter Gabor patch of main frequency 4 cpd and contrast 90%) surrounded by a square frame consisting of a black and white checkerboard (frame width of 8 pixels). The frame's contrast had an abrupt increase from 33% to 100% randomly between 300ms to 800ms from trial start and was used as visual exogenous cue. The target was then randomly presented within a 800 ms time window starting from cue onset and remained visible until subject's response. Subjects had to indicate the orientation of the target with the two fingers of the right hand by pressing two predefined keys on a standard keyboard. The subjects were asked to respond as fast as possible, but also to keep a high level of correct responses. Wrong responses as well as reaction times too fast (<150ms) or too slow (>1000ms) were discarded. For each subject at least 4 RTs per time bin were available and their mean used for obtaining the individual subject RTs versus time of target onset with respect to cue appearance. Time was sampled at 60 Hz. Written informed consent was obtained from each subject prior to the study and the experiment followed the tenets of the Declaration of Helsinki.

*Wavelets transformations and Fourier analysis*

The wavelets used throughout the work are Gabor (Morlet) wavelets defined in complex notation as:

$$G(t, f_G, \sigma_G) = (1/\sqrt{2\pi}\sigma_G) \exp\left(-\Delta t^2/(2\sigma_G^2) + i2\pi f_G \Delta t\right) \qquad \text{(Equ. 1)}$$

with $f_G$ its frequency, $\sigma_G$ its amplitude standard deviation, and $\Delta t$ the time deviation from the centre of the wavelet. Example wavelets of main frequencies 5, 10 and 15 Hz are shown in Figures 1 and 3 with their theoretical (red curve) and discrete points (dots) used to convolve it with the measured data. The "size" of the wavelet is defined as its length in time and is usually represented in multiples of $\sigma_G$; unless otherwise specified in the text, all wavelet values were $\sigma_G=0.85/f_G$; the wavelet window was $4\sigma_G$; $f_G$ is discretized in $1/T=1/0.8=1.25$ Hz steps where $T$ is the length of time window (here 800 ms). All Fourier analyses are carried with a square window in order to have its best sensitivity and not decrease sensitivity for signals close to the border of the time window.

*Theoretical signal model and simulations*
For the purpose of demonstrations and to assess the sensitivity of the single-condition method, simulations of a theoretical, fully controlled, experiment were performed. For this it is assumed that the measures were sampled at 60 Hz within a short time window of *T*=800 ms. A measurement *y(t)* of a theoretical oscillatory signal with frequency *f* and amplitude *A* that followed a Gaussian shape, with different rise ($\sigma_r$=20 ms) and decay ($\sigma_d$=100 ms) slopes, was simulated within the time window, and represented as:

$$y(t) = A(t) \times \cos(2\pi f(t - t_{peak}) + \phi) + \epsilon$$

$$A(t) = \begin{cases} A \exp\left(-(t - t_{peak})^2 / 2\sigma_r^2\right) & \text{for } t \leq t_{peak} \\ A \exp\left(-(t - t_{peak})^2 / 2\sigma_d^2\right) & \text{for } t \geq t_{peak} \end{cases} \quad \text{(Equ. 2)}$$

It's phase was random across repeated measures. Noise was an independent and identically distributed (i.i.d.) random variable. For the reaction times simulations the noise followed a log-normal distribution with mean 6.1 and standard deviation of 0.0964; for the simulations with white Gaussian noise, the mean was zero and standard deviation was one. The signal-to-noise ratio was defined as SNR=$A/\sigma_n$. It is supposed that *n*=20 repeated measures were performed (e.g. subjects). Out of the EEG data set all remaining time discretizations (simulations and data) had a sampling frequency of 60 Hz. The simulations for a given signal were as follows: (1) simulate n=20 measures of *y(t)* given the signal-to-noise ratio and peak's position; (2) compute their Fourier and TF decompositions; (3) create the reference distributions and perform the non-parametric test for each variable following the statistical procedures. For each SNR, peak time position, and frequency of signal 100 simulations of an experiment were carried and for each simulation the single-condition test and Fourier amplitude test were applied to test the presence or not of a signal (refer Figure 4).

*Statistical procedures*
The statistical procedures used for the two-condition comparison are based on creating Monte-Carlo permutation distributions that represent the null hypothesis of the statistics, as described in detail in other work (Maris, 2012; Maris and Oostenveld, 2007). For the TF non-parametric cluster-based test two reference distributions are necessary: (i) a distribution for each bin in the TF map obtained by shuffling the data between the two conditions $N_{shuffle}$ times; (ii) computation of a distribution for the cluster variable (size, sum of Z values…) that uses the result from the clusters of significant bins in each shuffled TF map and thus permits to create the reference distribution for clusters' statistics. For the single-condition procedure, one simulated measure $y_{1,p}(t)$ was obtained through Monte-Carlo time permutation by randomly shuffling the time points (subscript *p* indicates the p-th simulated measure), and repeating it $N_{shuffle}$ times for providing $N_{shuffle}$ versions of the original measures $y_1(t)$. Then, for each shuffled data $y_{1,p}(t)$ its TF representation was computed (and simple Fourier decomposition, wherever necessary; all $N_{shuffle}$=1000 unless otherwise specified). Then, the two reference distributions for *Stage-1* and *Stage-2* were created. The final cluster distribution was used for statistical testing to compute the probability to have obtained such a cluster in the data given the reference cluster distribution.

**Acknowledgments** C.B. was supported by the Deutsche Forschungsgemeinschaft (DFG) SFB 940 project B8. T.T. was partly supported by "the Fundamental Research Funds for the Central Universities" grant at USTC. This work was supported from the project "Sustainability for the National Institute of Mental Health", under grant number LO1611, with a financial support from the Ministry of Education, Youth and Sports of the Czech Republic under the NPU I program (D.K.).